# PWG-IDS: An Intrusion Detection Model for Solving Class Imbalance in IIoT Networks Using Generative Adversarial Networks

Lei Zhang[#], Shuaimin Jiang, Xiajiong Shen[#], Brij B. Gupta[*], *Senior Member, IEEE,* Zhihong Tian[*],
*Senior Member, IEEE*

*Abstract*—**With the continuous development of industrial IoT (IIoT) technology, network security is becoming more and more important. And intrusion detection is an important part of its security. However, since the amount of attack traffic is very small compared to normal traffic, this imbalance makes intrusion detection in it very difficult. To address this imbalance, an intrusion detection system called pretraining Wasserstein generative adversarial network intrusion detection system (PWG-IDS) is proposed in this paper. This system is divided into two main modules: 1) In this module, we introduce the pretraining mechanism in the Wasserstein generative adversarial network with gradient penalty (WGAN-GP) for the first time, firstly using the normal network traffic to train the WGAN-GP, and then inputting the imbalance data into the pre-trained WGAN-GP to retrain and generate the final required data. 2) Intrusion detection module: We use LightGBM as the classification algorithm to detect attack traffic in IIoT networks. The experimental results show that our proposed PWG-IDS outperforms other models, with F1-scores of 99% and 89% on the 2 datasets, respectively. And the pretraining mechanism we proposed can also be widely used in other GANs, providing a new way of thinking for the training of GANs.**

## I. INTRODUCTION

WITH the development of IIoT technology [1] and the introduction of strategies such as Made in China 2025 and German Industry 4.0, a wide variety of terminal manufacturing devices have become increasingly intelligent [2-6]. By connecting a variety of physical devices, It can perceive the real world at the level of vision, hearing, and touch. It can integrate information space with physical space, digitizes and network everything, realizes efficient information interaction between devices and devices, between devices and people, and between people and the real environment, and bring the informatization of the manufacturing industry to a higher level.

However, there is a huge amount of heterogeneous end devices in the IIoT. Although such highly distributed Internet devices can greatly expand the ability of the IIoT to sense the real world, this also poses a great security risk to IIoT security. Attackers can cause irreparable harm to individuals, enterprises, or countries by controlling end devices. For example, in 2015, attackers caused a major blackout in Ukraine by hacking into the circuit breakers of the country's electricity system and forcing a power outage on the grid system. In 2019, Ring, a home camera company, was exposed to a serious security vulnerability through which hackers could monitor users' homes and also expose their WiFi passwords. In June 2020, researchers in Germany found that SWARCO, the world's largest signal light controller manufacturing giant, has a security vulnerability that allows hackers to change the color of traffic lights at will through a debuggable open port of the traffic controller, which can cause serious traffic accidents.

Therefore, IIoT security has attracted the attention of many researchers. The IIoT network structure is divided into 3 layers: perception layer, network layer, and application layer. The security requirements of the sensing layer are mainly devoted to data security, such as preventing malicious node attacks, sample collection, and node data forgery damage. The security of the network layer is to prevent DoS attacks and ensure routing security. The application layer security is to meet the user privacy and access control. The current security mechanism for it mostly favors passive defense, whereas intrusion detection, an active protection method, can infer

This work was supported by the National Natural Science Foundation of China (Grant Nos. U20B2046, 61871140); Guangdong Province Universities and Colleges Pearl River Scholar Funded Scheme (2019); the Henan Key Laboratory of Big Data Analysis and Processing, School of Computer and Information Engineering, Henan University; Scientific and technological project of Henan Province (Grant No. 202102310340) , and Foundation of University Young Key Teacher of Henan Province(Grant No. 2019GGJS040, 2020GGJS027) and Key scientific research projects of colleges and universities in Henan Province(Grant No. 21A110005).

Lei Zhang is with Henan Key Laboratory of Big Data Analysis and Processing, Henan University, Kaifeng, 47500, China (e-mail: zhanglei@henu.edu.cn).

Shuaimin Jiang is with Institute of Data and Knowledge Engineering, Henan University, Kaifeng, 47500, China (e-mail: jsmxok@163.com).

Xiajiong Shen is with School of Computer and Information Engineering, Henan University, Kaifeng, 47500, China (e-mail: shenxj@henu.edu.cn).

Brij B. Gupta is with National Institute of Technology Kurukshetra, India, & Asia University, Taichung, Taiwan (e-mail: kpsannis@uom.edu.gr).

Zhihong Tian is with the Department of Cyberspace Insitute of Advanced Technology, GuangZhou University, GuangZhou, 510006 China (e-mail: tianzhihong@gzhu.edu.cn).

# Leizhang and Xiajiong Shen contributed equally to this manuscript.

* Correspondenc: Brij B. Gupta (e-mail: kpsa kpsannis@uom.edu.gr) and Zhihong Tian (e-mail: tianzhihong@gzhu.edu.cn).

suspicious attacks by analyzing network traffic data [7], which does not require too much modification of the original system, and can detect malicious behaviors when they are latent or occurring and improve effective help for further preventive measures, and this active protection technology is often more effective.

Currently, the research on intrusion detection is divided into two main models: IIoT device-centric intrusion detection research and collaboration-centric intrusion detection research. The IIoT device-centric intrusion detection research focuses on analyzing the potential threats of IIoT devices and designing intrusion detection algorithms oriented to a specific attack method with the help of techniques such as statistics and machine learning. For example, Ali Alheeti et al. designed and implemented an intelligent detection system for denial of service (DoS) attacks by analyzing the behavior of devices to determine whether they are under attack and then using fuzzy datasets to reduce the error and false alarm rates in the system [8]. However, IIoT device-centric intrusion detection models are often modeled and trained based on attack samples in a particular IIoT device, and their detection capabilities have limitations, are not well generalized, and cannot accurately sense the attack methods that have appeared in other devices, which causes it device-centric intrusion detection models to be ineffective in sensing a wide variety of potential threats in the IIoT. The rise of edge computing technologies in recent years has made collaboration-centric intrusion detection a hot topic of research for researchers. For example, Anup et al. proposed an intrusion detection model for mobile ad hoc networks (MANETs) [9], this model can detect the system parameters of MANETs in real-time and identify and detect various types of network attacks based on the changes of these parameters. Unfortunately, since the number of IIoT device intrusion samples is small and the communication between devices is slow, the collaboration-centric intrusion detection scheme is also difficult to cope with the complex and diverse threats in it. Researchers have, therefore, focused on cloud computing technology with massive data processing capability. Through cloud computing, each IIoT device can upload the acquired traffic features to the cloud center and use the powerful computing power of the cloud center combined with machine learning algorithms to build intrusion detection models. For example, Wu et al. proposed a knowledge transfer-based convolutional neural network model, which consists of two tandem convolutional neural networks, and can effectively improve the accuracy of intrusion detection by learning the underlying dataset and transferring the learned knowledge to the learning of the target dataset [10].

But the distribution of different types of attack traffic is extremely uneven in IIoT networks, and the number of attack samples in the minority class differs very much from the percentage of attack samples in the majority class. The class imbalance leading to the inability of the intrusion detection model to effectively detect network attacks in the minority class, which reduces the performance of the overall intrusion detection model.

To address this problem, this paper proposes PWG-IDS, which can expand the sample size of a minority class with a very small number of iterations, thus improving the performance of intrusion detection. The PWG-IDS consists of two modules, the data generation module and the detection module. The main component of the data generation module is the WGAN-GP [38], and the main component of the detection module is the LightGBM algorithm. The primary contributions of us are listed as follows:

- In order to solve the imbalance of minority class attack samples in IIoT networks, it is proposed for the first time to expand the number of minority class samples using WGAN-GP, and a filter is constructed to freely control the classes of the generated sample in the process of sample generation.

- Provides a new training method for GANs. Pioneering the use of pre-training mechanism to train WGAN-GP, all attack traffic can be regarded as a variant of normal traffic, so the probability distribution of normal traffic is first learned using WGAN-GP. Then, the learned probability distribution is used to generate a certain minority class, which can greatly reduce the number of iterations of WGAN-GP and generate more realistic samples. More importantly, this pre-training mechanism can also be used for other GANs.

- Based on the generation module, after comparing various algorithms, we select the best performance LightGBM algorithm as the intrusion detection module, which improves the performance of intrusion detection.

- Our comparative experiments are conducted using two representative intrusion detection datasets, and the results show that the PWG-IDS outperforms the state-of-the-art approach. In addition, this paper uses ablation studies to demonstrate the effectiveness of the pre-training mechanism in generating the minority class samples.

The rest of the paper is organized as follows: Section II details the principle and details of PWG-IDS proposed in this paper; Section III demonstrates experimentally the better performance of PWG-IDS compared with other advanced intrusion detection systems, and proves the effectiveness of our proposed pre-training mechanism through ablation studies; Section IV concludes the whole paper and proposes the future improvement directions of PWG-IDS.

## II. RELATED WORKS

### A. Intrusion Detection System (IDS)

IDS plays an essential role in cybersecurity, it improves the security of cyber by detecting attacks in machine learning, deep learning, or other algorithms.

In previous studies, many traditional machine learning algorithms have been used in IDS, such as Naive Bayesian [11,12], Random forest [13,14], SVM [15-17], and genetic algorithms [18].

In recent years, with the rise of deep learning techniques, more and more deep learning models have been used in intrusion detection and have obtained better performance than traditional machine learning. For example, Tang et al.

constructed an intrusion detection model based on multilayer perceptron (MLP) [19] and experimentally demonstrated that this model performs better than traditional machine learning algorithms such as SVM. Yin et al. considered that network traffic data can also be regarded as a kind of data related to time series, so they constructed an intrusion detection model based on recurrent neural network (RNN) and achieved better performance on the NSL- KDD, a dataset with relatively good results [20]. Gupta et al. constructed an intrusion detection system called LIO-IDS using long short-term memory (LSTM) network [21], this system has two components, the first component consists of LSTM to identify if it is attack traffic and the second component consists of an ensemble algorithm to classify the attack traffic. Basumalik et al., on the other hand, consider that traffic data possess some spatial information in addition to temporal information, and then, construct an intrusion detection model based on the Convolutional Neural Network (CNN) [22] and compare it with RNN, LSTM, experimentally prove that the use of CNN is more effective.

Unlike the above researchers who use machine learning or deep learning models to build intrusion detection systems, some researchers want to improve the performance of IDS through feature engineering. For example, Amiri et al. used mutual information algorithm for feature selection [23]. S Egea et al. proposed a new feature selection algorithm named Fast Based correlation Feature (FCBF) [24]. The idea of FCBF is to partition the feature space into several spaces of the same size, thus being able to distinguish sensor traffic data from other traffic data, which can effectively enhance the accuracy of the algorithms and reduce the execution time. Shafiq et al. proposed a new feature selection algorithm called CorrAUC based on packing technique [25]. Firstly, proposed a new feature selection metric CorrAUC based on the AUC metric, then designed a feature selection algorithm based on CorrAUC and used Shannon Entropy TOPSIS based on the bijective soft set technique to validate the features identified by CorrAUC, experimentally demonstrate that >96% results can be achieved used CorrAUC on the Bot-IoT dataset.

However, both ML and DL require a large amount of data for training, and both have very great difficulties in dealing with imbalanced data [40-51]. Researchers have proposed many algorithms to solve the class imbalance [52-54].

### B. Class Balancing Methods

The class imbalance problem has been viewed as a challenge in the field of ML and DL. Whenever there is a very large difference in the number of samples between multiple classes, it can be regarded as a class imbalance problem. Therefore, in IIoT network traffic with very unbalanced classes, IDS also needs to overcome this challenge [26,27].

Common techniques to solve class imbalance are random undersampling (RUS) and random oversampling techniques (ROS) [28]. Naturally, researchers first thought of using RUS and ROS to alleviate the class imbalance problem in intrusion detection [29,30]. However, the RUS technique loses some feature information, while ROS is very prone to overfitting, so Cieslak et al. used both RUS and ROS techniques to mitigate the class imbalance problem [31]. Similarly, to mitigate the drawbacks of RUS and ROS techniques, Tesfahun et al. applied Synthetic Minority Oversampling Technique (SMOTE) technique for intrusion detection [32]. Further, Yin et al. proposed a locally adaptive composite minority sampling algorithm called LA-SMOTE to alleviate the class imbalance problem of network traffic and to classify network traffic using GRU neural networks [33]. However, when the distribution between different categories is very similar, SMOTE will further blur the boundaries of different categories, which will instead lead to worse classification results. Therefore, Huang et al. instead proposed the Imbalanced Generative Adversarial Network (IGAN) model based on GAN to solve, the class imbalance problem in network traffic [34]. Firstly, using a filter to filter out the imbalanced traffic, and then, training the GAN individually based on the imbalanced traffic of a certain class to generate new traffic samples until all the imbalanced traffic is trained. Finally, the traffic generated by the GAN and the real traffic samples is trained together as a neural network to improve the classification of imbalanced classes.

Yet the number of unbalanced class samples is small and the convergence of GANs is difficult, GANs often take a very long time to complete the training, and even the generated sample distribution differs greatly from the real sample distribution. Therefore, this paper proposes the PWG-IDS based on the variant of GAN, WGAN-GP, and applies the pre-training mechanism to GANs network for the first time, which not only greatly improves the convergence speed of GANs network, but also through ablation study proves that the network traffic data generated by using the pre-training mechanism is closer to the real network traffic data.

### III. PWG-IDS IMPLEMENTATION

This section will introduce the PWG-IDS proposed in this paper in detail. As shown in Fig 1. PWG-IDS is composed of two modules: the data generation module and the intrusion detection module. The data generation module is implemented by WGAN-GP. We innovatively introduce the pre-training mechanism into GANs, which not only reduces the number of iterations of WGAN-GP, but also makes the generated data closer to the real data. And the convolutional layer is added to the WGAN-GP network to improve the characterization learning ability of the generator.

The intrusion detection module consists of feature extraction and classifier. In this paper, we use the Boruta [39] algorithm as the feature selection algorithm for the PWG-IDS model. The Boruta algorithm is a wrapper based on the random forest classification algorithm. The random forest classification algorithm is a relatively fast classification algorithm that can usually be implemented without adjusting the parameters and gives a numerical estimate of feature importance. Finally, LightGBM is used as a classifier for PWG-IDS to classify industrial IoT traffic and identify malicious traffic.

### A. Data Generator Module

We use WGAN-GP to implement the data generation module. WGAN-GP is developed from WGAN. Since WGAN uses weight clipping to satisfy the Lipschitz condition, it leads to the problems of training difficulty and slow convergence during the training process of WGAN. To solve this difficulty, WGAN-GP proposes to use gradient penalty to satisfy the Lipschitz condition.

Like the conventional GANs, the structure of WGAN-GP also contains a generation module (generator, G) and a discriminator module (discriminator, D). However, WGAN-GP sets an extra loss term in the discriminator to implement the gradient penalty, thus satisfying the Lipschitz condition. In other words, WGAN-GP sets an extra loss term in the discriminator so that the gradient does not exceed a K value. And the gradient penalty is not selected under the whole network, but only sampled between the true and false distributions, which can be formulated as:

$$W(\mathbb{P}_r, \mathbb{P}_g) \approx \max_{D} \left\{ \underset{x \sim \mathbb{P}_r}{\mathbb{E}}[D(x)] - \underset{\tilde{x} \sim \mathbb{P}_g}{\mathbb{E}}[D(\tilde{x})] \right. \\ \left. -\lambda \underset{\hat{x} \sim \mathbb{P}_w}{\mathbb{E}}[max(0, \|\nabla_{\tilde{x}} D(\tilde{x})\| - 1)] \right\} \quad (1)$$

Where $\mathbb{P}_r$ denotes the distribution of the real data, $\mathbb{P}_g$ denotes the distribution of the generated data. $W(\mathbb{P}_r, \mathbb{P}_g)$ indicates the degree of difference between the real and generated data distributions. $\underset{x \sim \mathbb{P}_r}{\mathbb{E}}, \underset{\tilde{x} \sim \mathbb{P}_g}{\mathbb{E}}$ denotes the expectation of the real data and the generated. $-\lambda \underset{\hat{x} \sim \mathbb{P}_w}{\mathbb{E}}$ denotes the gradient penalty term.

After adding the gradient penalty, the objective function of the final optimization of WGAN-GP is as follows:

$$L = \underset{\tilde{x} \sim \mathbb{P}_g}{\mathbb{E}}[D(\tilde{x})] - \underset{x \sim \mathbb{P}_r}{\mathbb{E}}[D(x)] + \lambda \underset{\tilde{x} \sim \mathbb{P}_w}{\mathbb{E}}\left[\left(\|\nabla_{\tilde{x}} D(\tilde{x})\|_2 - 1\right)^2\right] \quad (2)$$

$L$ indicates the final loss, where $\underset{\tilde{x} \sim \mathbb{P}_g}{\mathbb{E}}[D(\tilde{x})] - \underset{x \sim \mathbb{P}_r}{\mathbb{E}}[D(x)]$ indicates the final optimization function of WGAN. $\lambda \underset{\tilde{x} \sim \mathbb{P}_w}{\mathbb{E}}\left[\left(\|\nabla_{\tilde{x}} D(\tilde{x})\|_2 - 1\right)^2\right]$ is then the gradient penalty added by WGAN-GP.

In this paper, a pre-training mechanism and an imbalanced data filter are added to the WGAN-GP as the data generation module of the PWG-IDS model, and the structure of this module is shown in Fig 2. The data generation module contains three components: data filter, pre-training component, and imbalanced data generation component, where the network structure used for the pre-training component and the imbalanced data generation component is exactly the same, both of which are WAGN-GP. First, the data filter inputs the data labeled as normal traffic into the pre-training component. After the pre-training is completed, the pre-training component copies the trained network parameters to the unbalanced data generation component. At the same time, the data filter also inputs the imbalanced data into the data generation module. After the imbalanced data generation module is trained, the generated data is the data we ultimately need.

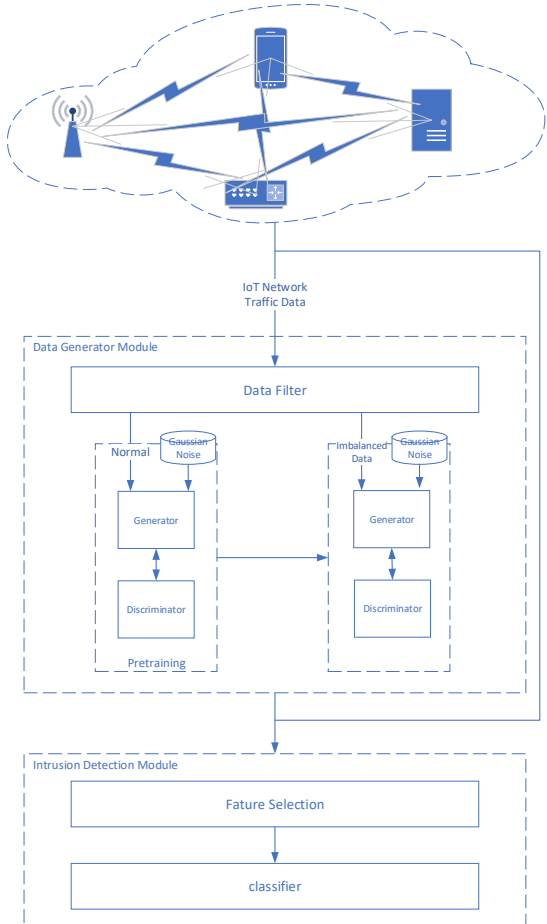

Fig. 1.  Architecture of PWG-IDS.

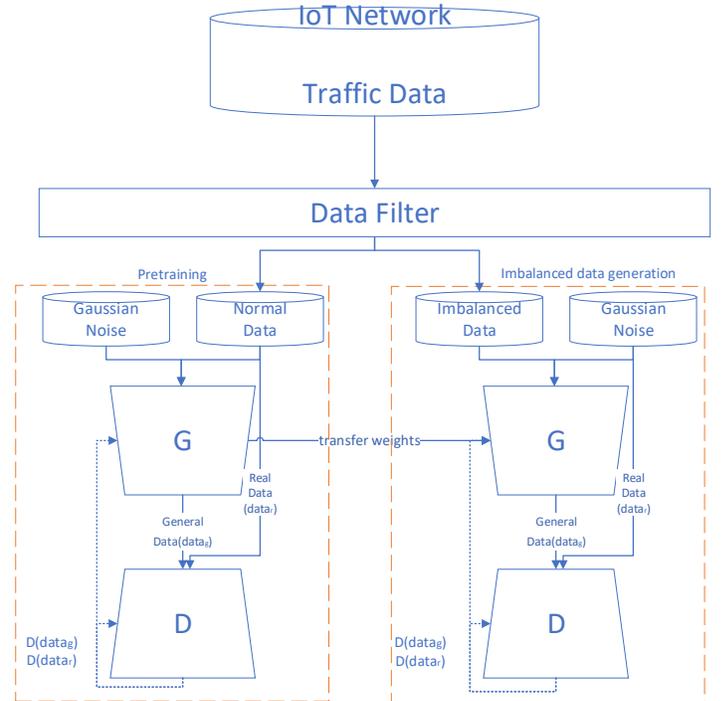

Fig. 2.  Data generator module of PWG-IDS.

a) Data Filter: Its input is $D_o = (X, y)$, indicates the traffic data of the IIoT, $X$ denotes the characteristics of the traffic, and $Y$ denotes the label of the data. The output is $D'_o = (X', y')$, then the filtering process of the filter is expressed as follows:

$$D'_o = \left\{ D'_o = (X', y') \mid D'_o \in D_o, y' \neq \underset{c_\tau \in \mathcal{C}}{\text{argmax}}(f_c(n_{c_\tau})) \right\} \quad (3)$$

$n_{c_\tau}$ indicates the number of data whose label is $c_\tau$, where

$C = (c_1, c_2, \cdots, c_\tau)$. $f_c(\cdot)$ is the imbalanced data selection function, and its formula is as follows:

$$f_c(n_{c_\tau}) = \begin{cases} n_{c_\tau}, & n_{c_\eta}/n_{c_\tau} \geq \gamma \\ 0, & n_{c_\eta}/n_{c_\tau} < \gamma \end{cases} \quad (4)$$

Where $n_{c_\eta} \in C$ denotes network traffic labeled as normal behavior and $n_{c_\tau} \in C$ denotes network traffic labeled as aggressive behavior. $\gamma$ indicates the imbalance rate. For example, if $\gamma = 10$ means when the number of attack traffic is less than or equal to 1/10 of the number of normal traffic, we consider it as imbalanced traffic which is the data that needs to be generated using the data generation module.

b) Pretraning and Imbalanced data generation Module: They are both WGAN-GP. Data Filter feeds normal data into the pre-training module because all attack traffic can be considered as a variant of normal traffic, so by using normal data for pre-training and copying the pre-trained network parameters into the data generation module, the number of iterations in the data generation module can be reduced and convergence can be accelerated.

### B. Intrusion Detection Module

In this paper, we use LightGBM as a classifier to detect attack traffic in IIoT networks. LightGBM is an efficient gradient boosting decision tree algorithm proposed by a Microsoft team [35], which belongs to the Gradient Boosting Decision Tree (GBDT) algorithm. GBDT is an additive model of a decision tree, which can be used The formula is expressed as:

$$f_m(x) = \sum_{m=1}^{M} T(x; \theta_m) \quad (5)$$

$T(x; \theta_m)$ is the decision tree; $\theta_m$ is the parameter of the decision tree; $M$ is the number of trees. The loss function is as follows:

$$L(y_i, f_m(x_i)) = \frac{1}{2}(y_i - f_m(x_i))^2 \quad (6)$$

where $y_i$ denotes the true value of the ith sample and $f_m(x_i)$ is the predicted value of the ith sample. Minimizing the loss function yields the optimal parameter $\hat{\theta}_m$:

$$\hat{\theta}_m = \arg\min \sum_{i=1}^{M} L(y_i; f_{m-1}(x) + T(x; \theta_m)) \quad (7)$$

Finally, the optimal GBDT model is constructed based on $\hat{\theta}_m$

However, the traditional GBDT algorithm takes up a lot of memory and time for training. LightGBM can take up less memory and reducing the training time by several times without losing prediction accuracy.

LightGBM uses two main methods to improve the training speed, 1) Gradient-based One-Side Sampling (GOSS), which selects data with larger gradients from the sample, thus improving the contribution to the computational information gain. 2) Exclusive Feature Bundling (EFB), which can merge certain features of the data and thus reduce the data dimensionality.

### C. Execution process of the model

a) Data filtering: As shown in Fig. 2, the traffic data are first passed through a data filter, and this filter will filter out the normal traffic data from the imbalanced traffic data. Then, normal traffic data are used as input for the pre-training module and imbalanced data is used as input for the data generation module.

b) Pre-training: The purpose of pre-training is to reduce the number of iterations when generating data, to speed up the convergence of the WGAN-GP, and to generate more realistic data. And in order to make the pre-trained network parameters available for the data generation module, the network structure of the pre-training module is designed to be exactly the same as that of the data generation module.

The pre-training module contains a generator $G$ and a discriminator $D$. The goal of $G$ is to make the distribution $\mathbb{P}_z(\hat{X}, \hat{y})$ of the generated data as close as possible to the distribution $\mathbb{P}_r(X, y)$ of the real data. And the goal of $D$ is to identify the difference between these two distributions as much as possible, then the process can be expressed by the formula:

$$\min_G \max_D V(D, G) = (\log D(X)) + \log(1 - D(G(\hat{X})) + \lambda \left( \left\| \nabla_{G(X)} D\left(G(\hat{X})\right) \right\|_2 - 1 \right)^2 \quad (8)$$

At each iteration, $G, D$ are optimized separately, i.e., G is first fixed to optimize $D$, and then $D$ is fixed to optimize $G$. Therefore, the optimization functions of $G, D$ can be expressed separately by the formula:

$$\max_{\theta_D} V(D) = \max_{\theta_D} \frac{1}{n} \sum_{i=1}^{n} \left( \log D(X_i) + \log(1 - D(G(\hat{X}_i)) + \lambda \left( \left\| \nabla_{\hat{X}_i} D(\hat{X}_i) \right\|_2 - 1 \right)^2 \right) \quad (9)$$

$\lambda$ denotes the weight of the penalty term in the gradient penalty, and according to the literature [38], the value of $\lambda$ is 10 by default. $\theta_D$ denotes the parameter in $D$. In this paper, the parameter is updated by the gradient descent method, which can be expressed by the following formula:

$$\theta_D = \nabla_{\theta_D} \left[ \frac{1}{m} \sum_{i=1}^{m} (\log D(X_i)) + \log(1 - D(\hat{X}_i)) + \lambda \left( \left\| \nabla_{\hat{X}_i} D(\hat{X}_i) \right\|_2 - 1 \right)^2 \right] \quad (10)$$

Equation (10) denotes the optimization function of D. Then, according to "(8), (9)", the optimization function of G can be obtained similarly as:

$$\min_{\theta_G} \hat{V}(G) = \min_{\theta_G} \frac{1}{n} \sum_{i=1}^{n} (\log(1 - D(G(\hat{X}_i))) \quad (11)$$

$\theta_G$ indicates the parameters in G, which update process is as follows:

$$\theta_G = \nabla_{\theta_G} [\frac{1}{m} \sum_{i=1}^{m} (\log(1 - D(\hat{X}_i)) \quad (12)$$

c) Imbalance data generation: The imbalance data generation module uses the network structure parameters $\theta_D, \theta_G$ of the pre-training module and the data filter to get imbalanced data as input. And use the imbalanced data to continue to train $\theta_D, \theta_G$. After adversarial training, $G$ can be used to generate class imbalanced data, pre-training and data generation module pseudo-code as algorithm 1.

d) Intrusion detection: $G$ is obtained according to step c), and then $G$ is used to generate class imbalance data, and both generated and real data are input to LightGBM for training, and the optimal classifier model is obtained after training is completed for detecting attack traffic in it.

| Algorithm 1: Imbalanced data generation |
|---|
| **Input:** $D_o = (X, y), D'_o = (\tilde{X}, \hat{y})$, We use default values of $lr = 0.0001, \beta_1 = 0.9, \beta_2 = 0.999, \lambda = 10$ |
| **Output:** generator |
| /* Process of pretrain */ |
| 1　 **while** $loss_G \geq 0.98$ and $loss_D \leq -0.98$ **do** |
| 　　　/* Training of the discriminator */ |
| 2　　 **for** t steps **do** |
| 3　　　 Sample $\{(X_i)\}_{i=1}^n$ as a batch form $\mathbb{P}_r(X, \ y)$ |
| 4　　　 Sample $\{Z_i\}_{i=1}^n$ as a batch from $\mathbb{P}_z(\tilde{X}, \hat{y})$ |
| 　　　　 $\tilde{X}_i \leftarrow G(Z_i)$ |
| 5　　　 $loss_D^i \leftarrow D(\tilde{X}_i) - D(X_i) + \lambda \left( \left\| \nabla_{\tilde{X}_i} D(\hat{X}_i) \right\|_2 - 1 \right)^2$ |
| 6　　　 $loss_D \leftarrow \frac{1}{n} \sum_{i=1}^n loss_D^i$ |
| 6　　　 $\theta_D \leftarrow \theta_D + lr \cdot Adam(\theta_D, \eta_{\theta_D}, \beta_1, \beta_2)$ |
| 7　　 **end for** |
| 　　　/* Training of the generator */ |
| 8　　 Sample $\{Z_i\}_{i=1}^n$ as a batch from $\mathbb{P}_z(\tilde{X}, \hat{y})$ |
| 9　　 $\theta_G \leftarrow \theta_G + lr \cdot Adam(\theta_G, \eta_{\theta_G}, \beta_1, \beta_2)$ |
| 10　 **end while** |
| 　　　/* Process of generating imbalanced data */ |
| 11　 **while** $loss_G \geq 0.99$ and $loss_D \leq -0.99$ **do** |
| 12　　 Use $\theta_D$ of the process of pre-train optimize discriminator |
| 13　　 Use $\theta_G$ of the process of pre-train optimize generator |
| 14　 **end while** |
| 15　 **return** G |

## IV. EXPERIMENT AND RESULT

In this section, experiments are conducted to evaluate the performance of the PWG-IDS proposed in this paper. First, the benchmark dataset, evaluation criteria, and execution environment used for the experiments are presented, and then, PWG-IDS is compared with the state-of-the-art model and class balancing methods, and the experiments show that the PWG-IDS outperforms the state-of-the-art methods. Then, the superiority of using the pre-training mechanism for training GANs networks is demonstrated through an ablation study.

### A. Benchmark Dataset

In this paper, NSL-KDD [36], CIS-IDS2018 [37], chosen as the benchmark dataset for evaluation, They are very classical datasets in the field of intrusion detection.

NSL-KDD is one of the first datasets to emerge in the field of intrusion detection, and it is an improvement of the KDD99 dataset, which each sample contains 41 features, where features 1-9 are basic features of network connections, features 10-22 are content-related traffic features, features 23-31 are time-related traffic features and features 31-41 are host-related traffic features.

CIS-IDS2018 is the most comprehensive and up-to-date dataset publicly available in the field of intrusion, created by the Canadian Institute for Cyber Security (CIC), and contains six different attack scenarios covering a wide range of currently known attacks. Each sample has 83 features, of which features

1-4 and 80-83 are basic network connectivity features, features 5-22 are network packet-related features, features 17-45 and 64-67 are network flow-related features, features 46-63 are content-related traffic features, and features 68-79 are traffic features related to generic targets.

The sample label distribution of the selected benchmark dataset in this paper is shown in Table I. It should be noted that due to the small number of samples in the NSL-KDD dataset, all of the NSL-KDD data are selected in this paper, while CIS-IDS2018 is only partially selected in this paper, but still has more samples than all of the NSL-KDD. Each dataset is divided by the ratio of the training set to the test set of 8:2.

TABLE I
DISTRIBUTION OF THE DATASETS.

| Dataset | Class | Samples | Imbalance ratio |
|---|---|---|---|
| NSL-KDD | Normal | 77054 | - |
| | DoS | 53385 | 1.443 |
| | Probe | 14077 | 5.474 |
| | R2L | 3749 | 20.553 |
| | U2R | 252 | 305.770 |
| CIS-IDS2018 | Benign | 80000 | - |
| | Bot | 22400 | 3.571 |
| | SSH-Bruteforce | 15200 | 5.263 |
| | FTP-Bruteforce | 15200 | 5.263 |
| | Infiltration | 12858 | 6.222 |
| | DDoS attack-HOIC | 50000 | 1.600 |
| | DDoS attack-LOIC-HTTP | 45600 | 1.754 |
| | DDoS attack-LOIC-UDP | 1384 | 57.803 |
| | DoS attacks-Hulk | 36800 | 2.174 |
| | DoS attacks-SlowHTTPTest | 10400 | 7.692 |
| | DoS attacks-GoldenEye | 3280 | 24.390 |
| | DoS attacks-Slowloris | 800 | 100.000 |
| | Brute Force-Web | 290 | 275.862 |
| | Brute Force-XSS | 121 | 661.157 |
| | SQL Injection | 43 | 1860.465 |

### B. Evaluation metrics

In order to evaluate PWG-IDS more accurately, we choose F1-Score and accuracy rate as the evaluation criteria, F1-Score is the summed average of the check accuracy rate and the check completeness rate, which is often used as the evaluation criteria for multi-classification problems. They are formulated as follows.

$$Accuracy = \frac{TP+TN}{TP+TN+FP+FN} \tag{13}$$

$$Precision = \frac{TP}{TP+FP} \tag{14}$$

$$Recall = \frac{TP}{TP+FN} \tag{15}$$

$$F1 - Scoer = 2 \times \frac{Precision \times Recall}{Precision + Recall} \tag{16}$$

Where TP, FP, TN, and FN indicate true positive, false positive, true negative, and plus negative, respectively.

## C. Execution Environment and Parameters

The experiments in this paper are done in Ubuntu, and the deep learning and machine learning frameworks used are Tensorflow and Sklearn, and the specific experimental environmental parameters are shown in Table II.

TABLE II
EXECUTION ENVIRONMENT

| Software / Hardware | Parameters |
|---|---|
| CPU | Inter Xeon E5v4 |
| GPU | Quadro P4000 |
| Memory | 64G |
| OS | Ubuntu 18.04 |
| DL Framework | Tensorflow 2.2 |
| ML Framework | Sklearn 0.24.2 |

The network structure of WGAN-GP used by the generation module is shown in Table III. The optimizer uses Adam, which has the parameters $lr = 0.00001, beta\_1 = 0.9, beta\_2 = 0.999, decay = 0$. The setting of batch size is determined according to the number of samples of the imbalance class, and the default is 64.

TABLE III
STRUCTURE OF WGAN-GP

| G-network | | | | |
|---|---|---|---|---|
| # | Layer | Size | Kernel | Activation |
| 1 | Fully-connected | Dimensions of input data | - | Leak ReLU |
| 2 | Convolutional | 64 | 1X3 | Leak ReLU |
| 3 | Convolutional | 32 | 1X3 | Leak ReLU |
| 4 | Convolutional | Dimensions of input data | 1X1 | - |
| D-network | | | | |
| 1 | Convolutional | 32 | 1X3 | Leak ReLU |
| 2 | Fully-connected | 64 | - | Leak ReLU |
| 3 | Dropout(rate=0.4) | - | - | - |
| 4 | Fully-connected | 1 | - | Tanh |

## D. Experimental Results

Since the traditional machine learning algorithms, such as naive Bayes, random deep forest, support vector machine, etc., generally do not perform as well as the GDBT algorithm, the common GDBT algorithm and deep neural network are chosen as baseline methods. In terms of class balancing, we have chosen SMOTEalgorithm, the experimental results are shown in Table IV.

From the above table, we can see that the GDBT methods achieves very good performance on the NSL-KDD dataset, but the accuracy of those methods is higher while the F1-Score is lower, and after mitigating the imbalance of the data by using the SMOTE algorithm, their F1-Score have improved significantly. It shows that GDBT methods cannot handle class imbalance data well by themselves. In contrast, the PWG-IDS proposed in this paper obtains the highest F1-Score on each dataset, proving that the PWG-IDS is very effective in dealing with the imbalance of the data.

## E. Ablation study

To demonstrate the effectiveness of the pre-training mechanism proposed in this paper, the PWG-IDS is ablated with CIS-IDS2018 as an example, and the impact of the pre-training mechanism on the generation module is first analyzed, and finally the impact of the pre-training mechanism on the performance of the whole model is analyzed.

Fig. 3 shows the number of iterations of the WGAN-GP network without pre-training, and Fig. 4 shows the number of iterations of the WGAN-GP network with pre-training, which clearly shows that the pre-training mechanism can greatly reduce the number of iterations of the WGAN-GP network, thus proving that the pre-training mechanism can effectively improve the convergence speed of the GNAs.

Also taking the CIS-IDS2018 dataset as an example, Fig. 5 shows the effect of the pre-training mechanism on the model Accuracy and F1-Score, and the following conclusions can be drawn from Fig. 3, Fig. 4 and 5: our proposed pre-training mechanism for GANs can reduce the convergence time of the GANs model without guaranteeing a reduction in model performance, thus improving the training speed. Moreover, the pre-training mechanism can also improve the performance of certain algorithms, such as XGBoost, CatBoost, and FNN.

TABLE IV
EXPERIMENTAL RESULTS

| Algorithm | NSL-KDD | CIC-IDS2018 | | |
|---|---|---|---|---|
| | Accuracy | Accuracy | Accuracy | F1-Score |
| LightGBM | 0.99 | 0.95 | 0.96 | 0.83 |
| XGBoost | 0.99 | 0.96 | 0.95 | 0.72 |
| CatBoost | 0.99 | 0.96 | 0.90 | 0.41 |
| FNN | 0.99 | 0.96 | 0.95 | 0.68 |
| LightGBM+SMOTE | 1 | 0.96 | 0.96 | 0.85 |
| XGBoost+SMOTE | 1 | 0.95 | 0.95 | 0.87 |
| CatBoost +SMOTE | 0.99 | 0.94 | 0.94 | 0.60 |
| FNN+SMOTE | 1 | 0.97 | 0.96 | 0.86 |
| **PWG-IDS** | **1** | **0.99** | **0.96** | **0.90** |

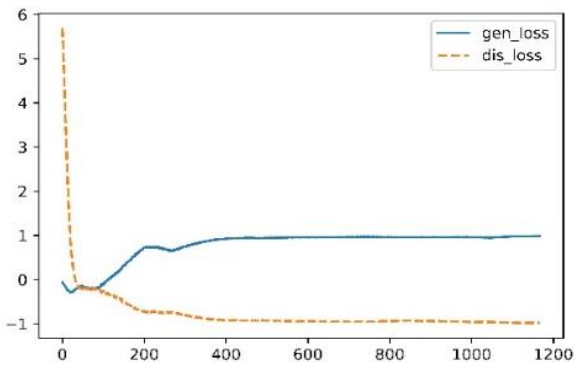

Fig. 3(a) Number of iterations required to generate sample of attack category DoS attacks-GoldenEye without pre-training

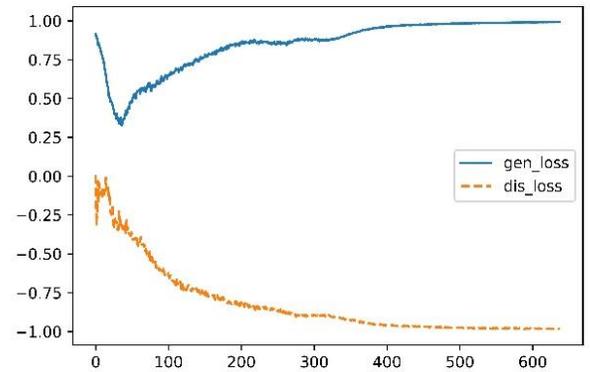

Fig. 4(a) Number of iterations required to generate sample of attack category DoS attacks-GoldenEye with pre-training

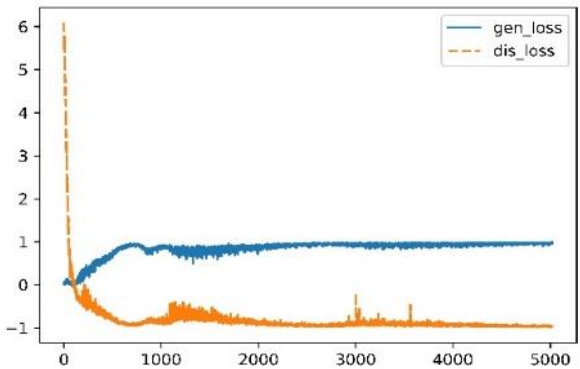

Fig. 3(b) Number of iterations required to generate sample of attack category SQL Injection without pre-training

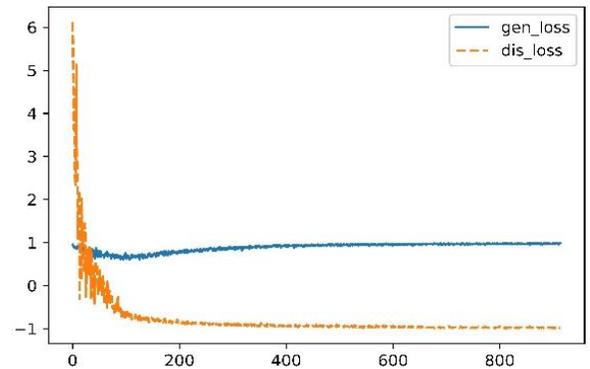

Fig. 4(b) Number of iterations required to generate sample of attack category SQL Injection without pre-training

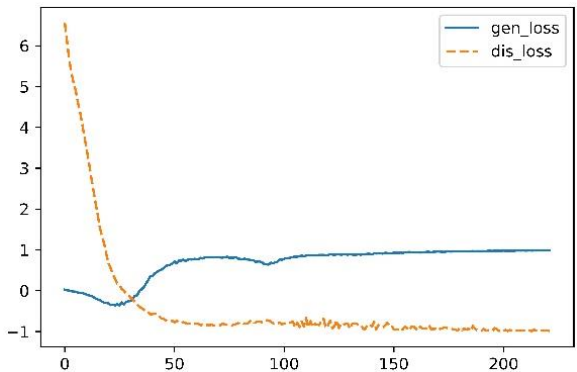

Fig. 3(c) Number of iterations required to generate sample of attack category DDoS attack-LOIC-UDP without pre-training

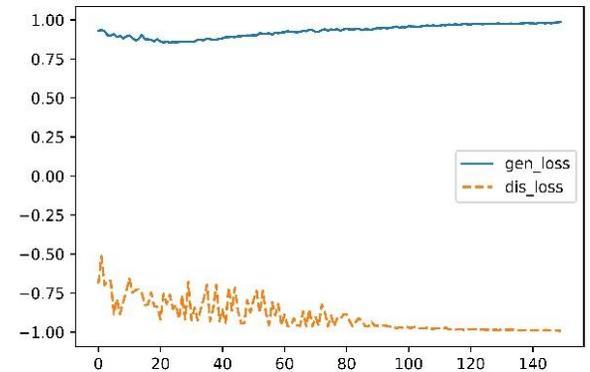

Fig. 4(c) Number of iterations required to generate sample of attack category DDoS attack-LOIC-UDP without pre-training

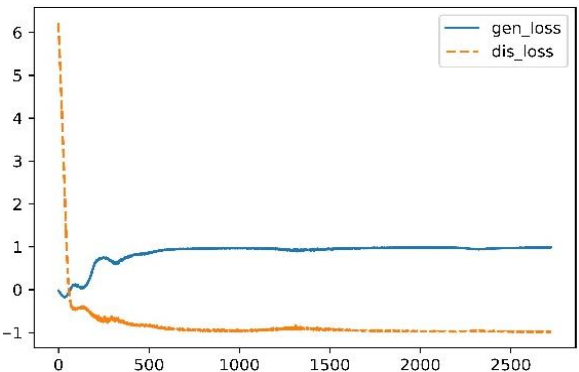

Fig. 3(d) Number of iterations required to generate sample of attack category DoS attacks-Slowloris without pre-trainin

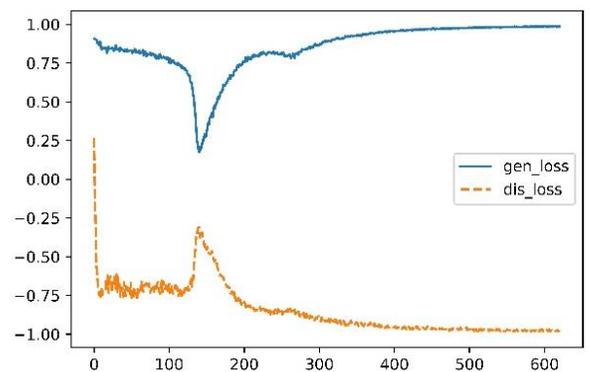

Fig. 4(d) Number of iterations required to generate sample of attack category DoS attacks-Slowloris without pre-trainin

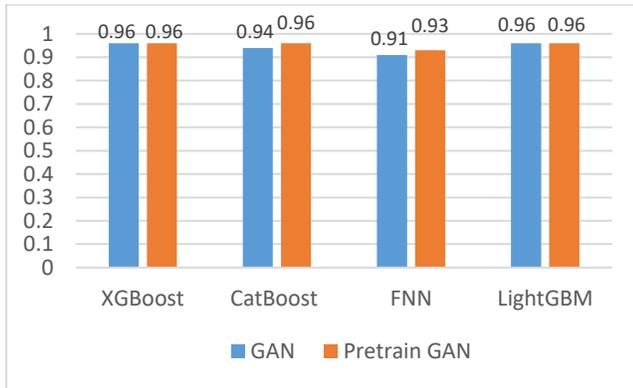

Fig. 5(a). Effect of pre-training mechanism on model Accuracy

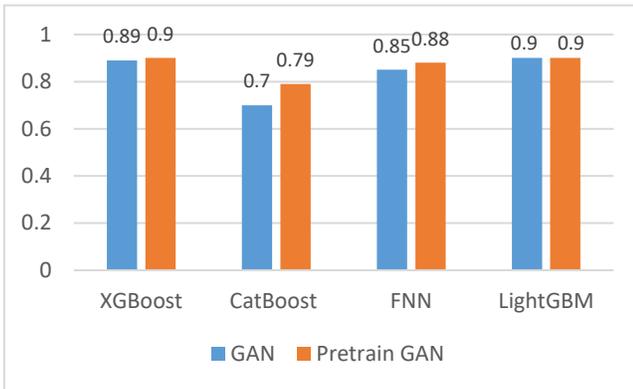

Fig. 5(b). Effect of pre-training mechanism on model F1-Scoer

## V. Conclusion

Intrusion detection is a hot issue in the field of IIoT security. With the exponential growth in the number of network devices, the imbalanced distribution of attack traffic poses great difficulties for IDS.

To solve this problem, PWG-IDS is proposed in this paper. First, we generate new network attack traffic samples for class imbalance traffic through the WGAN-GP, and in the process of generating samples, we also propose a pre-training mechanism for training GANs, which provides a new way of thinking for training GANs. We apply the pre-training network with WGAN-GP, which reduces the number of iterations of WGAN-GP, speeds up the convergence of GANs, and can generate more realistic data, effectively increasing the performance of the IDS. Then LightGBM is used as a classifier to detect attack traffic in IIoT networks. Finally, experiments are conducted using 2 typical intrusion detection datasets, and the experimental results show that the PWG-IDS proposed in this paper both outperform the current state-of-the-art methods.

However, the pre-training mechanism proposed in this paper requires separate pre-training for each different dataset, and, while using with pre-training cannot improve the performance of all classification algorithms. Therefore, in future work, we would like to increase the robustness and generalization capability of the pre-training mechanism.

## VI. References

[1] A. Al-Fuqaha, M. Guizani, M. Mohammadi, M. Aledhari, and M. Ayyash, "Internet of things: A survey on enabling technologies, protocols, and applications," IEEE Communications Surveys & Tutorials, vol. 17, no. 4, pp. 2347–2376, 2015.

[2] Y . Xiao, X. Du, J. Zhang, F. Hu, and S. Guizani, "Internet protocol television (IPTV): the killer application for the next-generation internet," IEEE Communications Magazine, vol. 45, no. 11, pp. 126–134, 2007.

[3] Z. Tian, S. Su, W. Shi, X. Du, M. Guizani, and X. Y u, "A data-driven method for future internet route decision modeling," Future Generation Computer Systems, vol. 95, pp. 212–220, 2019.

[4] Z. Tian, X. Gao, S. Su, J. Qiu, X. Du, and M. Guizani, "Evaluating reputation management schemes of internet of vehicles based on evolutionary game theory," IEEE Transactions on V ehicular Technology, 2019. 68(6): 5971-5980.

[5] Y . Xiao, V . K. Rayi, B. Sun, X. Du, F. Hu, and M. Galloway, "A survey of key management schemes in wireless sensor networks," Computer Communications, vol. 30, no. 11-12, pp. 2314–2341, 2007.

[6] X. Du and H.-H. Chen, "Security in wireless sensor networks," IEEE Wireless Communications, vol. 15, no. 4, pp. 60–66, 2008.

[7] Robert Mitchell and Ing-Ray Chen. 2014. A survey of intrusion detection techniques for cyber-physical systems. ACM Comput. Surv. 46, 4, Article 55 (April 2014), 29 pages. DOI https://doi.org/10.1145/2542049.

[8] K. M. Ali Alheeti, A. Gruebler and K. D. McDonald-Maier, "An intrusion detection system against malicious attacks on the communication network of driverless cars," 2015 12th Annual IEEE Consumer Communications and Networking Conference (CCNC), 2015, pp. 916-921, doi: 10.1109/CCNC.2015.7158098.

[9] Kumar Anup and P. Kumar Alampalayam Sathish "Intrusion detection and response model for mobile ad hoc networks"(2007).

[10] P. Wu, H. Guo and R. Buckland, "A Transfer Learning Approach for Network Intrusion Detection," 2019 IEEE 4th International Conference on Big Data Analytics (ICBDA), 2019, pp. 281-285, doi: 10.1109/ICBDA.2019.8713213.

[11] M. Panda , M.R. Patra , Network intrusion detection using Naive Bayes, Int. J. Comput. Sci. Netw. Secur. 7 (12) (2007) 258–263.

[12] N.B. Amor, S. Benferhat, Z. Elouedi, Naive Bayes vs decision trees in intrusion detection systems, in: Proceedings of the 2004 ACM Symposium on Applied Computing, in: SAC '04, ACM, Nicosia, Cyprus, 2004, pp 420–424, doi: 10.1145/967900.967989.

[13] J. Zhang, M. Zulkernine, and A. Haque, "Random-forests-based network intrusion detection systems," IEEE Trans. Syst., Man, Cybern. C, Appl. Rev., vol. 38, no. 5, pp. 649–659, Sep. 2008.

[14] N. Farnaaz and M. A. Jabbar, "Random forest modeling for network intrusion detection system," Procedia Comput. Sci., vol. 89, pp. 213–217, Jan. 2016.

[15] Y. Wang, J. Wong, A. Miner, Anomaly intrusion detection using one class SVM, in: Proceedings from the Fifth Annual IEEE SMC Information Assurance Workshop, 2004., IEEE, West Point, NY, USA, 2004, pp. 358–364, doi: 10.1109/IAW.2004.1437839.

[16] M.A.M. Hasan, M. Nasser, B. Pal, S. Ahmad, Support vector machine and random forest modeling for intrusion detection system (IDS), J. Intell. Learn. Syst. Appl. 6 (1) (2014) 45, doi: 10.4236/jilsa.2014.61005.

[17] H. Deng, Q.-A. Zeng, D.P. Agrawal, SVM-based intrusion detection system for wireless ad hoc networks, in: 2003 IEEE 58th Vehicular Technology Conference. VTC 2003-Fall (IEEE Cat. No. 03CH37484), volume 3, IEEE, Orlando, FL, USA, 2003, pp. 2147–2151, doi: 10.1109/VETECF.2003.1285404.

[18] Murray, S. N. , Walsh, B. P. , Kelliher, D. , & O'Sullivan, D. . (2014). Multi-variable optimization of thermal energy efficiency retrofitting of buildings using static modelling and genetic algorithms – a case study. Building and Environment, 75(MAY), 98-107.

[19] T. A. Tang, L. Mhamdi, D. McLernon, S. A. R. Zaidi and M. Ghogho, "Deep learning approach for Network Intrusion Detection in Software Defined Networking," 2016 International Conference on Wireless Networks and Mobile Communications (WINCOM), 2016, pp. 258-263, doi: 10.1109/WINCOM.2016.7777224.

[20] C. Yin, Y. Zhu, J. Fei and X. He, "A Deep Learning Approach for Intrusion Detection Using Recurrent Neural Networks," in IEEE Access, vol. 5, pp. 21954-21961, 2017, doi: 10.1109/ACCESS.2017.2762418.

[21] Neha Gupta, Vinita Jindal, Punam Bedi, LIO-IDS: Handling class imbalance using LSTM and improved one-vs-one technique in intrusion detection system, Computer Networks, Volume 192, 2021, 108076, https://doi.org/10.1016/j.comnet.2021.108076.

[22] Sagnik Basumallik, Rui Ma, Sara Eftekharnejad, Packet-data anomaly detection in PMU-based state estimator using convolutional neural network, International Journal of Electrical Power & Energy Systems, Volume 107, 2019, Pages 690-702, ISSN 0142-0615,

https://doi.org/10.1016/j.ijepes.2018.11.013.

[23] Fatemeh Amiri, MohammadMahdi Rezaei Yousefi, Caro Lucas, Azadeh Shakery, and Nasser Yazdani. 2011. Mutual information-based feature selection for intrusion detection systems. J. Netw. Comput. Appl. DOI: 10.1016/j.jnca.2011.01.002.

[24] S. Egea, A. R. Mañez, B. Carro, A. Sánchez-Esguevillas, and J. Lloret, "Intelligent iot traffic classification using novel search strategy for fast-based-correlation feature selection in industrial environments," IEEE Internet of Things Journal, vol. 5, no. 3, pp. 1616–1624, 2018.

[25] M. Shafiq, Z. Tian, A. K. Bashir, X. Du and M. Guizani, "CorrAUC: A Malicious Bot-IoT Traffic Detection Method in IoT Network Using Machine-Learning Techniques," in IEEE Internet of Things Journal, vol. 8, no. 5, pp. 3242-3254, 1 March1, 2021, doi: 10.1109/JIOT.2020.3002255.34 (4) (201 1) 1 184–1 199, doi: 10.1016/j.jnca.2011.01.002.

[26] S. Rodda, U.S.R. Erothi, Class imbalance problem in the network intrusion detection systems, in: 2016 International Conference on Electrical, Electronics, and Optimization Techniques (ICEEOT), IEEE, Chennai, India, 2016, pp. 2685–2688, doi: 10.1 109/ICEEOT.2016.7755181.

[27] V. Engen, J. Vincent, K. Phalp, Enhancing network based intrusion detection for imbalanced data, Int. J. Knowl. Based Intell. Eng. Syst. 12 (5–6) (2008) 357–367, doi: 10.3233/KES- 2008- 125- 605.

[28] A. Puri and M. K. Gupta, "Comparative Analysis of Resampling Techniques under Noisy Imbalanced Datasets," 2019 International Conference on Issues and Challenges in Intelligent Computing Techniques (ICICT), 2019, pp. 1-5, doi: 10.1109/ICICT46931.2019.8977650.

[29] Liwei Kuang and Mohammad Zulkernine. 2008. An anomaly intrusion detection method using the CSI-KNN algorithm. In Proceedings of the 2008 ACM symposium on Applied computing (SAC '08). Association for Computing Machinery, New York, NY, USA, 921–926. DOI: 10.1145/1363686.1363897.

[30] R. Abdulhammed, M. Faezipour, A. Abuzneid and A. AbuMallouh, "Deep and Machine Learning Approaches for Anomaly-Based Intrusion Detection of Imbalanced Network Traffic," in IEEE Sensors Letters, vol. 3, no. 1, pp. 1-4, Jan. 2019, Art no. 7101404, doi: 10.1109/LSENS.2018.2879990.

[31] D. A. Cieslak, N. V. Chawla and A. Striegel, "Combating imbalance in network intrusion datasets," 2006 IEEE International Conference on Granular Computing, 2006, pp. 732-737, doi: 10.1109/GRC.2006.1635905.

[32] A. Tesfahun and D. L. Bhaskari, "Intrusion Detection Using Random Forests Classifier with SMOTE and Feature Reduction," 2013 International Conference on Cloud & Ubiquitous Computing & Emerging Technologies, 2013, pp. 127-132, doi: 10.1109/CUBE.2013.31.

[33] Yan, B. , & Han, G. . (2018). La-gru: building combined intrusion detection model based on imbalanced learning and gated recurrent unit neural network. Security and Communication Networks,2018, 1-13.

[34] Shuokang Huang, Kai Lei, IGAN-IDS: An imbalanced generative adversarial network towards intrusion detection system in ad-hoc networks, Ad Hoc Networks, Volume 105, 2020, 102177, ISSN 1570-8705, https://doi.org/10.1016/j.adhoc.2020.102177.

[35] Ke Guolin, Meng Qi, Finley Thomas, Wang Taifeng, Chen Wei, Ma Weidong, Ye Qiwei and Liu Tie-Yan "LightGBM: A Highly Efficient Gradient Boosting Decision Tree."(2017).

[36] M. Tavallaee, E. Bagheri, W. Lu and A. A. Ghorbani, "A detailed analysis of the KDD CUP 99 data set," 2009 IEEE Symposium on Computational Intelligence for Security and Defense Applications, 2009, pp. 1-6, doi: 10.1109/CISDA.2009.5356528.

[37] I. Sharafaldin, A. H. Lashkari, and A. A. Ghorbani, "Toward generating a new intrusion detection dataset and intrusion traffic characterization," in Proc. 4th Int. Conf. Inf. Syst. Secur . Privacy, 2018, pp. 108–116.

[38] Gulrajani, I ., Ahmed, F . ,Arjovsky, M . , Dumoulin, V . , & Courville, A. . (2017). Improved Training of Wasserstein GANs.. 30, 5767-5777.

[39] Kursa, M. B., & Rudnicki, W. R. (2010). Feature selection with the Boruta package. J Stat Softw, 36(11), 1-13.

[40] Yu, C., Li, J., Li, X., Ren, X., et al (2018). Four-image encryption scheme based on quaternion Fresnel transform, chaos and computer generated hologram. Multimedia Tools and Applications, 77(4), 4585-4608.

[41] Alsmirat, M. A., Al-Alem, F., Al-Ayyoub, M., Jararweh, Y., et al. (2019). Impact of digital fingerprint image quality on the fingerprint recognition accuracy. Multimedia Tools and Applications, 78(3), 3649-3688.

[42] Wang, H., Li, Z., Li, Y., Gupta, B. B., & Choi, C. (2020). Visual saliency guided complex image retrieval. Pattern Recognition Letters, 130, 64-72.

[43] Al-Qerem, A., Alauthman, M., Almomani, A., & Gupta, B. B. (2020). IoT transaction processing through cooperative concurrency control on fog–cloud computing environment. Soft Computing, 24(8), 5695-5711.

[44] Gupta, B. B., & Quamara, M. (2020). An overview of Internet of Things (IoT): Architectural aspects, challenges, and protocols. Concurrency and Computation: Practice and Experience, 32(21), e4946.

[45] Bhushan, K., & Gupta, B. B. (2019). Distributed denial of service (DDoS) attack mitigation in software defined network (SDN)-based cloud computing environment. Journal of Ambient Intelligence and Humanized Computing, 10(5), 1985-1997.

[46] Adat, V., & Gupta, B. B. (2018). Security in Internet of Things: issues, challenges, taxonomy, and architecture. Telecommunication Systems, 67(3), 423-441.

[47] AlZu'bi, S., Shehab, M., Al-Ayyoub, M., Jararweh, Y., & Gupta, B. (2020). Parallel implementation for 3d medical volume fuzzy segmentation. Pattern Recognition Letters, 130, 312-318.

[48] Tewari, A., & Gupta, B. B. (2017). Cryptanalysis of a novel ultra-lightweight mutual authentication protocol for IoT devices using RFID tags. The Journal of Supercomputing, 73(3), 1085-1102.

[49] Al-Smadi, M., Qawasmeh, O., Al-Ayyoub, M., Jararweh, Y., & Gupta, B. (2018). Deep Recurrent neural network vs. support vector machine for aspect-based sentiment analysis of Arabic hotels' reviews. Journal of computational science, 27, 386-393.

[50] Stergiou, C., Psannis, K. E., Gupta, B. B., & Ishibashi, Y. (2018). Security, privacy & efficiency of sustainable cloud computing for big data & IoT. Sustainable Computing: Informatics and Systems, 19, 174-184.

[51] Tewari, A., & Gupta, B. B. (2020). Secure Timestamp-Based Mutual Authentication Protocol for IoT Devices Using RFID Tags. International Journal on Semantic Web and Information Systems (IJSWIS), 16(3), 20-34.

[52] J. Qiu, Z. Tian, C. Du, Q. Zuo, S. Su and B. Fang. A Survey on Access Control in the Age of Internet of Things. IEEE Internet of Things Journal. vol. 7, no. 6, pp. 4682-4696, June 2020, DOI: 10.1109/JIOT.2020.2969326.

[53] Y. Sun, Z. Tian, M. Li, S. Su, X. Du, and M. Guizani. Honeypot Identification in Softwarized Industrial Cyber-Physical Systems. IEEE Transactions on Industrial Informatics. vol. 17, no. 8, pp. 5542-5551, Aug. 2021. DOI:10.1109/TII.2020.3044576.

[54] Y. Wang, Z. Tian, Y. Sun, X. Du and N. Guizani. LocJury: An IBN-based Location Privacy Preserving Scheme for IoCV. IEEE Transactions on Intelligent Transportation Systems. 2021. 22(8): 5028-5037. DOI: 10.1109/TITS.2020.2970610.

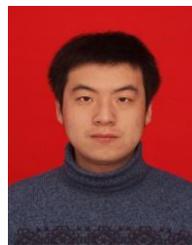**Lei Zhang** is an Associate Professor at Institute of Data and Knowledge Engineering, School of Computer and Information Engineering, Henan University, Henan Province, China. He received B.S. and M.S degree from Henan University Kaifeng China in 2003 and 2006, respectively, and received the Ph.D. degree in Computer Science and Technology from Harbin Institute of Technology, Harbin, China in 2015. His research interests include deep learning, computer networks and cyberspace security.

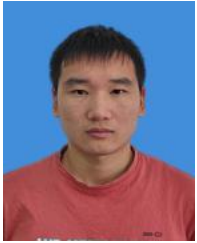

**Shuaimin Jiang** is now attending graduate school at Henan University School of Information Engineering. Its research interests are IoT information security and blockchain security.

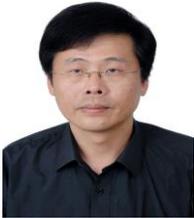

**Xiajion Shen Ph. D**., professor, PHD supervisor. Vice Dean of School of Computer and Information Engineering, Henan University. Executive Director of Henan Computer Society. Member of China Computer Federation. In 1986, he graduated from Wuhan University, Department of Computer Science and Technology, majoring in Computer System Architecture, with a Bachelor of Science degree. Since July 1986, he has been a faculty member in the Department of Computer Science and Technology, Henan University, during which he graduated from the School of Computer Engineering and Science, Shanghai University, with a Ph.D. degree in Control Theory and Control Engineering in 2006.

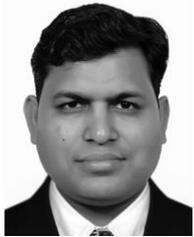

**Brij B. Gupta** (Senior Member, IEEE) received the Ph.D. degree in the area of information and cyber security from the Indian Institute of Technology Roorkee, India. He published more than 250 research papers in international journals and conferences of high repute. His biography was selected and published in the 30th Edition of Marquis Who's Who in the World, in 2012. He is working as Assistant Professor in the Department of Computer Engineering, National Institute of Technology Kurukshetra India & Asia University, Taichung, Taiwan. Moreover, he was also a Visiting Professor with the University of Murcia, Spain, from June 2018 to July 2018. He was a Visiting Professor with Temple University, USA (June 2019), and Staffordshire University, U.K. (July 2020). He is currently working as an Assistant Professor with the Department of Computer Engineering, National Institute of Technology Kurukshetra, India. His research interests include information security, cyber security, cloud computing, Web security, intrusion detection, and phishing.

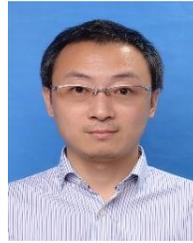

**Zhihong Tian** is currently a Professor, and Dean, with the Cyberspace Institute of Advanced Technology, Guangzhou University, Guangdong Province, China. Guangdong Province Universities and Colleges Pearl River Scholar (Distinguished Professor). He is also a part-time Professor at Carlton University, Ottawa, Canada. Previously, he served in different academic and administrative positions at the Harbin Institute of Technology. He has authored over 200 journal and conference papers in these areas. His research interests include computer networks and cyberspace security. His research has been supported in part by the National Natural Science Foundation of China, National Key research and Development Plan of China, National High-tech R&D Program of China (863 Program), and National Basic Research Program of China (973 Program). He also served as a member, Chair, and General Chair of a number of international conferences. He is a Senior Member of the China Computer Federation, and a Member of IEEE.